\documentclass[oneside, reqno, 10pt]{amsart}

\usepackage{amsmath,amssymb}
\usepackage[dvips]{geometry}
\usepackage{color}
\usepackage{graphicx,epsfig,subfigure,pgf, epstopdf}

\numberwithin{equation}{section}
\theoremstyle{plain}                
\newtheorem{theorem}{Theorem}[section]
\newtheorem{lemma}[theorem]{Lemma}
\newtheorem{proposition}[theorem]{Proposition}
\newtheorem{corollary}[theorem]{Corollary}
\newtheorem{definition}[theorem]{Definition}

\newtheorem{assumption}[theorem]{Assumption}
\newtheorem{remark}[theorem]{Remark}

\newenvironment{Proof}{\noindent\textsc{Proof.}}{\mbox{ } \hfill $\Box$ \vspace{2mm}}

\def\be{\begin{equation*}}
\def\ee{\end{equation*}}

\DeclareMathOperator*\esssup{esssup}

\newcommand{\tot}{\tfrac{1}{2}} 
\newcommand{\abs}[1]{\left| #1 \right|} 
\newcommand{\set}[1]{\left\{#1\right\}} 
\newcommand{\sets}[2]{\set{#1\,:\,#2}} 

\newcommand{\norm}[1]{{||#1||}} 

\newcommand{\ft}[2]{#1\dots#2}
\renewcommand{\ft}[2]{#1,\dots,#2}

\newcommand{\mapstto}{\twoheadrightarrow\!\!\!\!\!\!\mapsto}

\providecommand{\R}{} \renewcommand{\R}{{\mathbb R}}

\newcommand{\e}{{\mathfrak{E}}}
\newcommand{\Ff}{{\mathcal{F}}}

\newcommand{\N}{{\mathbb N}}
\newcommand{\PP}{{\mathbb P}}

\newcommand{\FF}{{\mathcal F}}
\newcommand{\Pp}{{\mathcal P}}

\renewcommand{\AA}{{\mathcal A}}
\newcommand{\bsy}[1]{\boldsymbol{#1}}
\newcommand{\bS}{\boldsymbol{\Sigma}}
\newcommand{\bmu}{\boldsymbol{\mu}}

\newcommand{\bone}{\bsy{1}}
\newcommand{\bsi}{\bsy{\sigma}}
\newcommand{\balpha}{\bsy{\alpha}}
\newcommand{\bzeta}{\bsy{\zeta}}
\newcommand{\bze}{\bzeta}
\newcommand{\bzt}{\bzeta'}

\newcommand{\bpi}{\bsy{\pi}}

\newcommand{\Xze}{X^{\bzeta}}

\newcommand{\optbz}{\bzeta^*}
\newcommand{\Xoze}{X^{\optbz}}
\renewcommand{\bS}{\bsy{S}}
\newcommand{\bW}{\bsy{W}}

\newcommand{\cprf}[1]{\{ #1 \}_{t\in [0,T]}}

\newcommand{\vnot}[2]{ (#1)_{i=\ft{1}{#2}} }
\newcommand{\vnotr}[2]{ (#1)_{j=\ft{1}{#2}} }

\newcommand{\mnot}[3]{(#1)_{i=\ft{1}{#2}}^{j=\ft{1}{#3}}}

\newcommand{\zmu}{\zeta_{\bmu}}
\newcommand{\zsigma}{\zeta_{\bsi}}

\newcommand{\E}{\mathbb{E}}

\begin{document}
\renewcommand{\baselinestretch}{1.1}\normalsize

\begin{center}
  \huge\sc CRRA Utility Maximization under Risk Constraints\footnote{
Santiago Moreno--Bromberg gratefully acknowledges financial support from  the Deutsche
Forschungsgemeinschaft through the SFB 649 ``Economic Risk'' and from the Alexander von Humboldt Foundation via a research fellowship.\\
Traian A. Pirvu is grateful to NSERC through grant  5-36700 and MITACS through grant 5-26761.
\\ Anthony R\'eveillac is grateful to the Deutsche
Forschungsgemeinschaft Research center MATHEON for financial support. The authors are very grateful to Jianing Zhang for his guidance concerning the numerical simulations.}
\end{center}

\medskip

\begin{center}
\today
\end{center}

\medskip

\begin{center}
\begin{tabular}{ccc}
{Santiago Moreno--Bromberg} & \hspace{1cm} & {Traian A.~Pirvu} \\
Insitut f\"{u}r Banking und Finance & & Mathematics and Statistics Department \\
Universit\"{a}t Z\"{u}rich   & & McMaster University \\
Plattenstr. 32, 8032 Z\"{u}rich & & Hamilton, ON \\
Switzerland & &  Canada\\
{\tt santiago.moreno@bf.uzh.ch} & & {\tt tpirvu@math.mcmaster.ca} \\
\end{tabular}
\end{center}
\vspace{.4cm}

\begin{center}
Anthony R\'eveillac  \\
Universit\'e Paris Dauphine\\
CEREMADE UMR CNRS 7534\\
Place du Mar\'echal De Lattre De Tassigny\\
75775 Paris Cedex 16\\
France\\
{\tt anthony.reveillac@ceremade.dauphine.fr}
\end{center}
\vspace{.6cm}

\begin{center}
{\bf Abstract}
\end{center}
This paper studies the problem of optimal investment
with CRRA (constant, relative risk aversion) preferences, subject to
dynamic risk constraints on trading strategies. The market model considered is continuous in time and incomplete. the prices of financial assets are modeled by It\^{o} processes. The dynamic risk constraints, which are time and state dependent,
are generated by risk measures. Optimal trading strategies are characterized by a quadratic BSDE.
Within the class of \textit{time consistent distortion risk measures}, a three--fund separation result is
established. Numerical results emphasize the effects of imposing risk constraints on trading.

\vspace{0.2in}

\begin{center}
\textsc{Preliminary - Comments Welcome}
\end{center}\vspace{0.2in}

\begin{flushleft}
{\bf JEL classification: }{G10}\\
{\bf Mathematics Subject Classification (2000): }
{91B30, 60H30, 60G44}\\
{\bf Keywords:} BSDE, CRRA preferences, constrained utility maximization, correspondences, risk measures.
\end{flushleft}

\pagebreak
\pagestyle{plain}

\setcounter{equation}{0}
\section{Introduction}

In this paper we consider the problem of  a utility--maximizing agent, whose preferences are of of constant relative risk aversion (CRRA)  type, and whose trading strategies are subject to risk constraints.
We work on a continuous--time, stochastic model with randomness being driven by Brownian noise. The market
is incomplete and consists of several traded assets whose prices follow It\^{o} processes.

\vspace{.1in}
In practice, managers set risk limits on the strategies executed by their traders. In fact, the mechanisms used to control risk are more complex: financial institution have specialized internal departments in charge of risk assessments. On top of that there are external regulatory institutions to whom financial institutions must periodically report their risk exposure. It is natural, therefore, to study the portfolio problem with risk constraints, which has received a great deal of scrutiny lately. A very well known paper in this direction is \cite{CviKar92}. The authors employ convex duality to characterize the optimal constrained portfolio. A more recent paper in the same direction is \cite{HIM05}. Here the optimal constrained portfolio is characterized by a quadratic BSDE, which renders the method more amenable to numerical treatment. In these two (by now classical) papers the risk constraints are imposed via abstract convex (closed) sets. Lately, a line of research has been developed where the risk--constraint sets are specified employing a specific risk measure, e.g. VaR (Value at Risk). In the following we provide a brief overview of the related literature.

\vspace{.2in}
{\bf{Existing Research:}}
A risk measure that is commonly used by both practitioners and academics is VaR. Despite its success, VaR has as drawbacks  {not being subadditive and not recognizing the accumulation of risk}. This encouraged researchers to develop other risk measures, e.g. TVaR (Tail Value at Risk). The works on optimal investment with risk constraints generated by VaR, TVaR (or other risk measures)
 split into two categories, which depend on whether or not the risk assessment is performed in a static or a dynamic fashion. Let us briefly touch on the first category.
 The seminal paper is \cite{BasSha01}, where the optimal dynamic portfolio and wealth-consumption policies
of utility maximizing investors who use VaR to control their risk exposure is analyzed. In a complete--market,
It\^o-processes framework, VaR is computed in a static manner (the authors compute the VaR of the final wealth only). An interesting finding is that VaR limits, when applied only at maturity, may actually increase risk.  {One way to overcome this problem is to} consider a risk measure that is based on the risk--neutral expectation of  loss - the Limited Expected Loss (LEL). In \cite{EmmKluKor01}  a model with Capital--at--Risk (a version of VaR) limits, in the Black--Scholes--Samuelson framework is presented. The authors assume that portfolio proportions are held constant during the whole investment period, which makes the problem static. \cite{DmiLarLiWar10} extends \cite{EmmKluKor01} from constant to deterministic parameters.
In a market model with constant parameters, \cite{GabWun09} extends \cite{BasSha01} to cover the case of bounded expected loss. In a general, continuous--time Financial market model, \cite{GunWeb07} considers the portfolio problem under a downside risk constraint measured by an abstract convex risk measure.
\cite{Kup09} extends \cite{EmmKluKor01} by imposing a uniform (in time) risk constraint.

\vspace{.1in}
In the category of dynamic risk measurements we recall the seminal paper \cite{CuoHeIss08}. Following the financial industry practice, the VaR
(or some other risk measure) is computed (and dynamically re--evaluated) using a time window (2 weeks in practice) over which the trading strategies
are assumed to be held constant for the purpose of risk measurement. The finding of the authors is that dynamic VaR and TVaR constraints reduce
 the investment (proportion wise) in the risky asset. \cite{LeiVanTro06} studies the impact of VaR constraint on  equilibrium prices and the relationship
 with the leverage effect. \cite{BerCumKou02} shows that, in equilibrium, VaR reduces market volatility. \cite{Pri10} finds that risk constraints may give rise to equilibrium asset pricing bubbles. Among others, \cite{AtkPap05}, \cite{Pir07}, and \cite{Yiu04} analyze  the problem of investment and consumption subject to dynamic VaR constraints. \cite{PirZit09} considers maximizing the growth rate of the portfolio in the context of dynamic VaR, TVaR and LEL constraints. In a complete market model, \cite{Sas10} uses a martingale method to study the optimal investment under dynamic risk constraints
 and partial information.

\vspace{.2in}
{\bf{Our Contribution:}}
This paper extends the risk measurements introduced by \cite{CuoHeIss08} by considering a relatively
general class of risk measures (we only require them to be Carath\'eodory maps, and this class is rich enough to include many convex and coherent risk measures). The corresponding risk--constraint sets  {arising} from such risk measures, and applied to the trading strategies, are time and state dependent. Moreover, they satisfy some important measurability properties.

\vspace{.1in}
We employ the method developed in \cite{HIM05} in order to find the optimal trading strategies subject to
the risk constraints. The main difference is that, unlike \cite{HIM05}, our constraint sets are time dependent, {which renders}
the methodology developed in  \cite{HIM05} not directly applicable within our context. The difficulty stems from establishing the measurability of the BSDE's driver (the BSDE which characterizes the optimal trading
strategy). This is done by means of the Measurable Maximum Theorem and the Kuratowski--Ryll--Nardzewski Selection Theorem. After this step is achieved we apply results from \cite{Morlais09} to get existence of solutions to the BSDE, which in turn yields the optimal trading strategy.

\vspace{.1in}
We then restrict our risk measures to the class of \textit{time consistent distortion risk measures}. By doing so we observe that
the risk constraints have a particular structure: they are compact sets (for a fixed time and state)
and depend on two statistics (portfolio return and variance). This leads to a three--fund separation result. More precisely, an investor subject to
regulatory constraints will invest her wealth into three--funds: a savings account and two index funds.
One index fund is a mix of the stocks with weights given by the Merton proportion. This index fund is
related to market risk and most of the portfolio separation results refer to it. The second index is related
to volatility risk. In a market with non--random drift and volatility the second index is absent. Thus, the second
index can be explained by the demand of hedging volatility risk.

\vspace{.1in}
Numerical results shed light into the structure of the optimal trading strategy. More precisely, using recent results concerning numerical methods for quadratic growth BSDEs, we present in Section \ref{section:numerics} some numerical examples for value--at--risk, tail--value--at--risk and limited expected loss. Our simulations clearly exhibit the effect of the risk constraint on the optimal strategy and on the associated value function. More precisely
from the plots we see that risk constraints reduce the gambling of the risky assets.

\vspace{.2in}
The paper is organized as follows: In Section \ref{section:Model} we
introduce the basic model, the risk measures
and the corresponding risk constraints. Section \ref{sec:analysis}
presents measurability properties of the candidate optimal trading strategy
and its characterization via a quadratic BSDE. In Section \ref{sec:VaR}, \textit{time consistent distortion risk measures} are considered. A three--fund separation result is obtained within this context. Numerical results are presented in Section \ref{section:numerics}. The paper ends with an appendix that contains some technical results.

\setcounter{equation}{0}

\section{Model Description and Problem Formulation}\label{section:Model}

\subsection{The Financial Market}\label{ssec:FinMarkt}
 Our model of a financial market, based on a filtered probability space
$(\Omega,\FF,\cprf{\FF_t},\PP)$ that satisfies the usual conditions,
consists of $n+1$ assets. The first one, $\cprf{S_0(t)}$, is a {\em
  riskless bond} with a strictly positive, constant interest rate $r>0$. The
remaining $n$ assets are {\em stocks}, and their prices are modeled by an
$n$--dimensional It\^o--process
$\cprf{\bS(t)}=\cprf{\vnot{S_i(t)}{n}}$. Their  dynamics are given by the following stochastic differential
equations, in which $\cprf{\bW(t)}=\cprf{\vnot{W_i(t)}{m}}$ is a
$m$--dimensional standard Brownian motion:
\be%
\label{equ:SDSs-for-stocks}
    \nonumber
\left.
\begin{aligned}
  dS_0(t)& = S_0(t) r \, dt \\
  dS_i(t)& =S_i(t)\Big( \alpha_i(t)\, dt+\sum_{j=1}^m \sigma_{ij}(t)\,
  dW_j(t) \Big),\ i=\ft{1}{n},
    \end{aligned}
\right\}, t\in [0,T],
\ee
where the $\R^n$--valued process $\cprf{\balpha(t)}=\cprf{\vnot{\alpha_i(t)}{n}}$ is the
{\em mean rate of return}, and $\cprf{\bsi(t)}$
$=\cprf{\mnot{\sigma_{ij}(t)}{n}{m}}\in\R^{n\times m}$ is the
{\em variance--covariance} process. In order for the equations
\eqref{equ:SDSs-for-stocks} to admit unique strong solutions, we impose the
following regularity conditions on the coefficient processes
$\balpha(t)$ and $\bsi(t)$:
\begin{assumption}\label{ass:basic-regularity}
 All the components of the processes $\cprf{\balpha(t)}$ and $\cprf{\bsi(t)}$ are
predictable, and

\be
    \nonumber
    \begin{split}
      \sum_{i=1}^n \int_0^t \abs{\alpha_i(u)}\, du+\sum_{i=1}^n
      \sum_{j=1}^m \int_0^t \sigma_{ij}(u)^2\, du<\infty,\,\text{ for
        all $t\in [0,\infty)$, $\PP$-a.s.}
    \end{split}
\ee

\end{assumption}
\noindent To ease the exposition, we introduce the following notation: for an integrable
  $\R^m$-valued process $\bsy{\gamma}(t)=\vnot{\gamma_i(t)}{n}$, and a
  sufficiently regular $\R^m$--valued process
  $\bpi(t)=\vnotr{\pi_j(t)}{m}$ we write

\be
    \nonumber
    \begin{split}
      \int_0^t \bsy{\gamma}(u)\, du:= \sum_{i=1}^n \int_0^t
      \gamma_i(u)\, dt,\quad
      \int_0^t \bpi(t)\, d\bW(t):= \sum_{j=1}^m \int_0^t
      \pi_j(t)\, dW_j(t).
    \end{split}
\ee
Further, we impose the following condition on the variance--covariance
process $\bsi(t):$

\begin{assumption}
\label{ass:independent-rows}
The matrix $\bsi(t)$ has   independent rows for all $t\in [0,\infty)$ {almost--surely}.
\end{assumption}

\noindent This assumption makes it impossible for different stocks to have the same
diffusion structure. Otherwise, the market would either allow for
 arbitrage opportunities or  redundant assets would exist.
 {As a consequence of Assumption \ref{ass:independent-rows} we have} that $n\leq m$ - the
number of risky assets does not exceed the number of ``sources of
uncertainty''. Also, the inverse $(\bsi(t)\bsi(t)')^{-1}$ is easily
seen to exist. The equation
\be
    \begin{split}
\bsi(t) \bsi(t)' \bze_M(t)= \bmu(t)
    \end{split}
\ee
uniquely defines a predictable
stochastic process $\cprf{\bze_M(t)}$, named the
{\em Merton--proportion process}, where $\cprf{\bmu(t)}=\cprf{\vnot{\mu_i(t)}{n}}$, with
$\mu_i(t)=\alpha_i(t)-r$ for $i=\ft{1}{n}$. At this point we make another assumption on
the market coefficients:

\begin{assumption}\label{ass:boundmprx}
We assume that
\be
\E\left[ \exp{\Big(\int_0^T \norm{\bze_M(t) \bsi(u)}^2\, du\Big)}\right]<\infty,
\ee
and the stochastic process $\bsy{\sigma'}\bsy(\bsi \bsy{\sigma'})^{-1}\bsi$ is uniformly bounded.
In addition, we assume that there exists a constant $c>0$ such that
\be
\norm{\bsy{\sigma'}\bsy(\bsi \bsy{\sigma'})^{-1}\bsi(t) \bmu(t)} \leq c, \; \forall t \in [0,T], \; \PP-a.s..
\ee
\end{assumption}

\subsection{Trading strategies and wealth}\label{ssec:TradStrat}

Let $\Pp$ denote the predictable $\sigma$--algebra on $[0, T]\times\Omega.$
The control variables are the proportions of
current wealth  the investor invests in the assets. More precisely, we
have the following formal definition:
\begin{definition}\label{def:portfolio-proportions}
An $\R^n$--valued stochastic process
$\cprf{\bze(t)}$ $=\cprf{\vnot{\zeta_i(t)}{n}}$ is called an {\em
  admissible portfolio--proportion  process} if
it is predictable (i.e. $\Pp$-measurable) and {it}
satisfies
\begin{equation}%
\label{equ:regularity-zeta}
    \nonumber
    \begin{split}
      \E\Big[\int_0^t \abs{\bzt(u)(\balpha(u)-r\bone)}\, du+\int_0^t
      \norm{\bzt(t) \bsi(u)}^2\, du \Big]<\infty,\quad\text{for all}\quad
      t\in [0,\infty).
    \end{split}
\end{equation}
Here $\bzt(t)$ denotes the transpose of $\bzeta(t),$ $\bone=(1,\dots,1)'$ is a $n$--dimensional column
vector all of whose coordinates are equal to $1$, and
$\norm{\bsy{x}}$ is the standard Euclidean norm. The set of admissible {strategies will be} denoted by $\AA$.
\end{definition}

\noindent Given a portfolio--proportion process $\bze(t)$, we interpret its $n$
coordinates as the proportions of the current wealth $\Xze(t)$
invested in each of the $n$ stocks. In order for the portfolio to be self--financing, the
remaining wealth $\Xze(t)(1-\sum_{i=1}^n \bzeta_i(t))$ is assumed to be
invested in the riskless bond $S_0(t)$.  {If} this quantity
is negative, we are effectively borrowing at the rate $r>0$.
No short--selling restrictions are imposed, hence the
proportions $\bzeta_i(t)$ are allowed to be negative, and they are not a priori bounded.
{The} equation governing the evolution of the total wealth $\cprf{\Xze(t)}$
of the investor using the portfolio--proportion process
$\cprf{\bze(t)}$ is given by

\begin{equation}%
\label{equ:wealth-one}
    \nonumber
    \begin{split}
d\Xze(t)&= \Xze(t) \Big(
\bzt(t) \balpha(t)\, dt+\bzt(t)\bsi(t)\, d\bW(t) \Big)
+\Big(1-\bzt(t)\bone\Big) \Xze(t)r \, dt\\
&= \Xze(t) \Big( (r+\bzt(t) \bmu(t))\, dt+\bzt(t)\bsi(t)\,
d\bW(t)
\Big).
    \end{split}
\end{equation}
We recall that $\cprf{\bmu(t)}=\cprf{\vnot{\mu_i(t)}{n}}$, with
$\mu_i(t)=\alpha_i(t)-r$ for $i=\ft{1}{n}$, is the vector of {\em
  excess rates of return}.  Under {the}
regularity conditions \eqref{equ:regularity-zeta} imposed on $\bze(t),$
Equation~\eqref{equ:wealth-one} admits a unique strong solution given by

\begin{equation}%
\label{equ:wealth-two}
    \nonumber
    \begin{split}
      \Xze(t)=X(0)\exp\Big\{ \int_0^t
        \Big(r+\bzt(u)\bmu(u)-\tot\norm{\bzt(u)\bsi(u)}^2\Big)\, du+
        \int_0^t \bzt(u) \bsi(u)\, d\bW(u) \Big\}.
    \end{split}
\end{equation}
The initial wealth $\Xze(0)=X(0)\in (0,\infty)$
 {is considered to be} exogenously given. As a consequence of Assumption \ref{ass:boundmprx}, {and} using Expression \eqref{equ:regularity-zeta}, a strategy $\bzeta$ is admissible if and only if it is a predictable process such that
\begin{equation}\label{tri}
\E\Big[\int_0^T \norm{\bzt(u) \bsi(u)}^2\, du\Big]<\infty.
\end{equation}
Indeed  we have
\begin{equation}
    \nonumber
    \begin{split}
 \bzt (u)\bmu (u)= (\bsi' (u) \bze (u))' (\bsi'(u) \bze_M (u))\leq
 \norm{\bzt \bsi(u)} \,\norm{\bze_M'(u) \bsi (u)},
    \end{split}
\end{equation}
by the Cauchy--Buniakowski--Schwarz inequality. Thus, inequality \eqref{tri} follows from
Assumption \ref{ass:boundmprx}, Expression \eqref{equ:regularity-zeta} and  the Cauchy--Buniakowski--Schwarz inequality.

The expression appearing inside the first integral in
\eqref{equ:wealth-two} above will be given its own notation; the quadratic function
$\tilde{Q}:\R^2\to\R$ is defined as

\be
    \begin{split}
      \tilde{Q}(\zmu,\zsigma):= r+\zmu-\tot \zsigma^2,
    \end{split}
\ee
It is also useful to define the random field $Q:\Omega\times
[0,\infty)\times \R^n\to\R$

\be
    \begin{split}
      Q(t,\bze):=\tilde{Q}(\bzt \bmu(t), \norm{\bzt\bsi(t)}).
    \end{split}
\ee
\noindent It is clear from {Expression} \eqref{equ:wealth-two} that
the evolution of wealth process $\Xze(t)$ depends on the $\R^n$-dimensional process
$\bze(t)$ only through two ``sufficient statistics'', namely
\begin{equation}%
\label{equ:portfolio-quantities}
    \nonumber
    \begin{split}
 \zmu(t):= \bzt(t)\bmu(t),\text{ and }\zsigma(t):=
\norm{\bzt(t)\bsi(t)}.
    \end{split}
\end{equation}
These will be referred to in the sequel as
{\em portfolio rate of return} and
{\em portfolio volatility}, respectively.

\subsection{Projected  distribution of wealth}\label{sub:projected-wealth}

For the purposes of risk measurement, it is common practice to use an approximation
of the distribution of the investor's wealth at a future date.
Given the current time $t\geq 0$, and a length $\tau>0$
of the measurement horizon $[t, t+\tau),$
the {\em projected distribution} of the wealth from trading are
calculated under the simplifying assumptions that
\begin{enumerate}
\item the proportions of the wealth $\{\bze(s)\}_{s\in
    [t, t+\tau)}$  invested in various
  securities, as well as
\item the market coefficients $\{\balpha(s)\}_{s\in [t, t+\tau)}$
  and $\{\bsi(s)\}_{s\in [t, t+\tau)}$
\end{enumerate}
stay constant and equal
to their present values throughout the time interval $[t, t+\tau)$.
The wealth Equations \eqref{equ:wealth-one} and
\eqref{equ:wealth-two} yield that the {\em projected wealth loss} is
- conditionally on $\FF_{t}$ -
distributed as $L=L(X(t),
\zmu(t),\zsigma(t))$, where the law of $L(x,\zmu,\zsigma)$
is the one of
\begin{equation}
\label{equ:law-of-L}
   \nonumber
    \begin{split}
x \Big(1-\exp (Y(\zmu,\zsigma))\Big).
    \end{split}
\end{equation}
Here $Y(\zmu,\zsigma)$ is a normal random variable
with mean $\tilde{Q}(\zmu,\zsigma)\tau$ and
standard deviation $\sqrt{\tau}\zsigma$. The quantities $\zmu(t)$
and $\zsigma(t)$ are the portfolio rate of return and volatility,
defined in Equation~\eqref{equ:portfolio-quantities}. In the upcoming sections we turn our focus to risk measurements associated to the {\em relative projected wealth gain}, which will be defined as the distribution of the quantity
\be
\frac{\Xze(t+\tau-){-}\Xze(t)}{\Xze(t)}.
\ee
This is not a technical requirement, and the method developed in Sections~\ref{subsecc:generalRM} to~\ref{sec:analysis} still holds for risk measurements in absolute terms. The economic implications, however, may be stark, and the definition of the risk constraints below would require a certain recursive structure. The latter in the sense that admissibility (risk--wise) at time $t$ will depend on the choice of the strategy at all previous times. We elaborate further on this in Remark~\ref{rem:absolute-vs-relative}. The measurement horizon $\tau$ and the market coefficients will play the role of ``global variables''.

\subsection{The risk constraints}\label{subsecc:generalRM}

In this section we  {introduce} the risk constraints that will be imposed on the trading strategies. We keep the presentation as general as possible and make only sufficient assumptions on the risk measures. These allow us to show existence (and in some cases uniqueness) of optimal, constrained trading strategies. We begin by making precise how the risk of a given strategy {is measured}.\\

\noindent
Let us define the gain over time interval $[t, t+\tau]$ by $\Delta_{\tau}X^{\bzeta}_t := X^{\bzeta}_{t+\tau-}-X^{\bzeta}_t,$ and let $(\rho_t)_{t\in[0,T]}$ be a family of maps $\rho_t$ with
\be
\rho_t:\mathcal{C}_t \subset L^2(\mathcal{F}_T, \PP) \to L^2(\mathcal{F}_t, \PP),
\ee
where
\be
\mathcal{C}_t:=\Big\{\Delta_{\tau}X^{\bzeta}_t/X^{\bzeta}_t\,\Big|\; \bze \;{\text{ is an admissible strategy}}\Big\}.
\ee
Notice that for all $t\in (0, T],$ we have that $\mathcal{C}_t\subset L^2(\mathcal{F}_T, \PP).$ We also define $\mathcal{C}_0:= L^2(\mathcal{F}_T, \PP).$ For a given admissible $(\tilde\bzeta(s))_{s\in [0, t)}$ and $\bzeta\in\R^n$ we define the strategy $\overline{\bzeta}:\Omega\times[0, t+\tau)\to\R^n$ as  $\overline{\bzeta}(s)=\tilde\bzeta(s)$ for $s<t$ and $\overline{\bzeta}(s)=\bzeta$ for $t\le s<t+\tau.$ By definition of the wealth process we obtain that $X^{\overline{\bzeta}}_t=X^{\tilde\bzeta}_{t-},$ moreover (under the assumptions made in Section~\ref{sub:projected-wealth}) the quantity
$\Delta_{\tau}X^{\overline{\bzeta}}_t/X^{\overline{\bzeta}}_t$ depends exclusively on $\bzeta,$ and not on $\tilde\bzeta.$
In order to establish the risk constraints,  we define the acceptance sets

\begin{equation}\label{eq:Admissible-sets}
\nonumber
\AA_t^{\rho,\tilde\bzeta}(\omega):=\left\{\bzeta\in\R^n\,\Big|\, \rho_t\Big(\frac{\Delta_{\tau}X^{\overline{\bzeta}}_t}{X^{\overline{\bzeta}}_t(\omega)}\Big)(\omega)\le K_t(\omega)\right\}\quad t\in [0, T],
\end{equation}
where $K_t$ is a real--valued, exogenous, predictable process  {that satisfies} $K_t\geq \rho_t(0)$ for all $t$ in $[0,T]$, $P$--almost surely. Notice that $\bzeta=0$ is in the constraint set. We observe that by construction, the sets $\AA_t^{\rho,\tilde\bzeta}$ are independent of $\tilde\bzeta,$ and we shall simply write $\AA_t^{\rho}.$ In analogous fashion we will slightly abuse notation and write $\Delta_{\tau}X^{\bzeta}_t/X^{\bzeta}_t$ for $\Delta_{\tau}X^{\overline{\bzeta}}_t/X^{\overline{\bzeta}}_t.$ It follows from Equation~\eqref{equ:wealth-one} that in fact
\be
\frac{\Delta_{\tau}X^{\bzeta}_t}{X^{\bzeta}_t} = \e(\bzeta, t)-1,
\ee
where
\be
\e(\bzeta, t):=\exp\left\{\int_t^{t+\tau}\big(r+\zmu(u)-\frac{1}{2}\zsigma(u)^2\big)du+\int_t^{t+\tau}\bzeta(u)\bsi(u) dW(u)\right\}.
\ee
Hence, the expressions for the sets of constraints $\AA^{\rho}_t$ may be rewritten as
\be
\AA^{\rho}_t(\omega)=\left\{\bzeta\in\R^n\,\Big|\, \rho_t(\e(\bzeta,t)-1)(\omega)\le K_t(\omega)\right\}.
\ee
Moreover, under the assumption that $\bmu,$ $\bsi$ and $\bzeta$ remain (for the purpose of risk assessment) constant over $[t, t+\tau),$ we may write
\be
\e(\bzeta, t):=\exp\Big\{r\tau\Big\}\cdot\exp\Big\{\tau\big(\zmu-\frac{1}{2}\zsigma^2\big)\Big\}\cdot\exp\Big\{\bzeta \bsi\Delta_{\tau}W_t\Big\},
\ee
and we shall denote by $\e_1(\bzeta, t)$ and $\e_2(\bzeta, t)$ the second and third factors of $\e(\bzeta, t),$ respectively.

\noindent {We} make the following assumption on the family $(\rho_t)_{t \in [0,T]}:$
\begin{assumption}\label{ass:rho}
The family of maps
\be
\rho_t:\mathcal{C}_t \subset L^{2}({\mathcal F}_T, \PP) \to L^{2}({\mathcal F}_t, \PP)
\ee
satisfies that the mapping
\be(\bzeta,(\omega,t)) \mapsto \rho_t(\e(\bzeta, t)-1)(\omega)
\ee
is a Carath\'eodory function; that is, for every $(\omega,t)$ in $\Omega \times [0,T]$, the map $\bzeta  \mapsto \rho_t(\e(\bzeta, t)-1)(\omega)$ is continuous and for every $\bzeta$ in $R^n$  the map $(\omega,t)\mapsto \rho_t(\e(\bzeta, t)-1)(\omega)$ is $\Pp$--measurable.
\end{assumption}
\noindent
An example of a family $(\rho_t)_{t\in [0,T]}$ that satisfies Assumption \ref{ass:rho} is the following: Let $l:\R\to\R$ be a convex, non--decreasing continuous and non--constant function\footnote{Such functions are usually referred to as ``loss functionals''.} with $|l(-\infty)|<+\infty$. Assume that the filtration $\{\mathcal{F}_t\}_{t\in [0,T]}$ is generated by the Brownian motion $\cprf{\bW(t)}$ and that $\sigma_{i,j}(t):=\sigma_{i,j}(t,W_t)$ and $\mu(t):=\mu(t,W_t)$ where $\sigma_{i,j}$ and $\mu$ are deterministic Borelian functions. We set
\be
\rho_t(-\e(\zeta,t)-1) = \E[l(\exp(r (\tau+\zeta x \frac12 \|\zeta y\|^2)+x y \Delta_\tau W_0))]_{x=\mu(t,W_t), y=\sigma(t,W_t)}
\ee
so that $\rho_t(-\e(\zeta,t)-1)=\mathbb{E}[l(-\e(\zeta,t)-1)\vert \mathcal{F}_t]$, $\PP$--almost surely.
Then the family $(\rho_t)_{t\in [0,T]}$ satisfies Assumption \ref{ass:rho}. Indeed, fix $\bsy{a}$ in $\R^n$ and
let $\bzeta$ in $\R^n$. Then, by monotonicity of the exponential and $l$ we have that:
$$ l(-\infty) \leq l(-\e(\bzeta,t)+1) \leq l(1).$$
Hence, Lebesgue's Dominated Convergence Theorem implies that:
\be
\lim_{\bzeta \to \bsy{a}} \rho_t(-\e(\zeta,t)-1) = \rho_t(-\e(\bsy{a},t)-1), \quad \forall t\in [0,T].
\ee
Finally, since the filtration we consider is the Brownian filtration, the stochastic process $(\rho_t(\e(\bzeta, t)-1))_{t\in [0,T]}$ is predictable.

\begin{remark}\label{rem:absolute-vs-relative} If we were to consider risk constraints based not on the relative projected wealth
loss, but only on the quantities $\Delta_{\tau}X^{\bzeta}_t,$ then the acceptance sets defined in Expression~\ref{eq:Admissible-sets} would depend on $(\bzeta(s))_{s\in[0, t)}.$ More precisely, the set of risk--admissible strategies would be
\be
\AA:=\Big\{\bzeta = (\bzeta(s))_{s\in [0, T]}\,\big|\, \bzeta\;{\text{is admissible and}}\; \bzeta(t)\in\AA^{\rho, \bzeta{\mbox{\small\bf{1}}}_{[0, t)}}_t\Big\}
\ee
In the case where $\rho_t$ is a $\Ff_{t_{-}}$--coherent family, i.e. if $\rho_r(XY)=X\,\rho_t(Y)$ for all $X\in \Ff_{t-},$ then risk constraints in absolute terms are generated by inequalities of the form
\be
X^{\overline{\bzeta}}_t\rho_t\Big(\frac{\Delta_{\tau}X^{\overline{\bzeta}}_t}{X^{\overline{\bzeta}}_t}\Big)\le K_t.
\ee
This follows from the fact that the wealth level at time $t$ is a $\Ff_{t-}$--measurable random variable. The structure then reverts to that of risk constraints in relative terms, except for a redefinition of the risk bound as $\tilde{K}_t(\omega):= K_t(\omega)/X_t(\omega).$ Notice that if $K_t\equiv K\in\R_+,$ then $\tilde{K}_t$ would be a decreasing function of wealth. In other words, highly capitalized investors would face more stringent constraints. This could lend an approach to dealing with the too--big--to--fail problem, and could be further tweaked by allowing ${K}_t$ to depend on the state of nature. It is, however, beyond the scope of this paper to discuss such policy--making issues, and we shall stick to the relative--measures--of--risk framework.
\end{remark}

\begin{remark}
Note that $(\rho_t)_{t\in [0,T]}$ is not \textit{stricto sensu} a dynamic risk measure, since every $\rho_t$ is a priori not defined on the whole space $L^{2}({\mathcal F}_T, P)$. As we we have seen in the previous lines, defining the risk of {every} random variable in $L^{2}({\mathcal F}_T, P)$ is not {relevant} for us, since we only need to evaluate the risk of the very specific random variables $\Delta_{\tau}X^{\bzeta}_t$.
\end{remark}

\subsection{The optimization problem}

We finish the section by formulating our central problem.
Given a choice of a dynamic risk measure $\rho$ satisfying Assumption \ref{ass:rho}
and a final date $T,$ we are
searching for a portfolio--proportion process $\optbz(t)\in\AA^{\rho}_t$ which maximizes the $p-$CRRA utility $U_{p}(x)=\frac{x^{p}}{p}, p<1,$
of the final wealth among all the portfolios satisfying the same constraint. {In other words},
 for all $t\in [0, \infty)$ {and} $\bze(t)\in\AA^{\rho}_t=\left\{\bzeta\in\R^n\,\Big|\, \rho_t(\e(\bzeta,t)-1)\le K_t\right\}$

\begin{equation}\label{eq:opti}
    \begin{split}
\mathbb{E}\big[U_{p}(\Xoze(T))\big]\geq
\mathbb{E}\big[U_{p}(\Xze(T))\big].
    \end{split}
\end{equation}
This problem has the following economic motivation: Risk managers limit the risk exposure of their
traders by imposing risk constraints on their strategies. This can be regarded as an external risk
management mechanism. In our model this is represented by the risk measures. On the other hand, traders
have their own attitudes towards risk, which are reflected by the risk aversion of the CRRA utility. However,
$p\in[0,1)$ is known to reflect a risk seeking attitude of the trader. The risk manager cannot constraint
the trader's risk preferences. In order to deal with this, risk constraints on the trader's strategies
must be imposed.
\section{Analysis}\label{sec:analysis}

In this section we prove the existence of an optimal investment strategy. For simplicity we consider
the case $p\in(0,1)$ (analogous arguments apply with minor modifications to $p<0$). In order to do so, we make use of the powerful theory of backward stochastic differential equations (BSDEs). Let
\be
\AA^\rho := \big\{\bzeta=(\bzeta(t))_{t\in[0,T]} \in \AA \,\big|\, \bzeta(t) \in \AA^{\rho}_t, \; \forall t\in [0,T] \big\},
\ee
where $\AA$ is the set of admissible strategies in the sense of Definition \ref{def:portfolio-proportions}.
We recall that we consider the maximization problem
\be
\max_{\bzeta \in \AA^\rho} \mathbb{E}(U_{p}(\Xze(T))).
\ee
By means of Equation \eqref{equ:wealth-two} we may write
\be
U_p(X^{\bze}(t))=U_{p}(X(0))\exp\left( \int_0^t p\Big(r+\zmu(u)-\tot\zsigma(u)^2\Big)\, du+
        \int_0^t p\,\bzt(u) \bsi(u)\, d\bW(u) \right).
\ee
In analogous fashion as done in~\cite{HIM05}, let us introduce the auxiliary process
\be
R^{\bze}(t):= U_{p}(X(0))\exp\left(Y(t)+\int_0^t
        p\Big(r+\zmu(u)-\tot\zsigma(u)^2\Big)\, du+\int_0^t p\,\bzt(u) \bsi(u)\, d\bW(u) \right),
\ee
where $(Y,Z)$ is a solution to the BSDE

\begin{equation}\label{BSDE}
Y(t)=0 - \int_{t}^{T}Z(u)d\bW(u) - \int_{t}^{T}h(u,Z(u))du,\qquad t\in[0,T].
\end{equation}
{The} function $h(t,z)$ {should be chosen} in such a way that

\begin{itemize}
\item[a)] the process $R^{\bze}$ is a supermartingale, $R^{\bze}(T)=U_p(\Xze(T))$ and $R^{\bze}(0)=  \frac{(X(0))^p}{p}$ for every $\bze\in \AA^\rho$,
\item[b)] there exists at least one element $\bze^*$ in $\AA^\rho$ such that $R^{\bze^\ast}$ is a martingale.
\end{itemize}
We shall verify ex--post that the function $h(t, z)$ in question satisfies the measurability and growth conditions required to guarantee existence of solutions to Equation~\eqref{BSDE}. Before going further we explain why {achieving this would provide} a solution to Problem \eqref{eq:opti}. If we {were} able to construct such a family of processes $R^{\bze},$ then we {would obtain} that $\bze^*$ is an optimal strategy for Problem \eqref{eq:opti} with initial capital $X(0)>0$ independent of $\bze$. Indeed let $\bze$ any element of $\AA^\rho$, then using (a) and (b) {we have}
$$ \mathbb{E}(U_{p}(\Xze(T))=\mathbb{E}(R^{\bze}(T)) \leq  R^{\bze}(0) = \frac{(X(0))^p}{p} = \mathbb{E}(R^{\bze^\ast}(T)). $$
This method is known as the \textit{martingale optimality principle}. Let us now perform a multiplicative decomposition
of $R^{\bze}$ into martingale and an increasing process. Given a continuous process $M,$ we denote by $\mathcal{E}(M)$ its stochastic exponential:
\be
\mathcal{E}(M(t)):=\exp\left(M(t)-\frac{1}{2}\langle M\rangle_{t}\right),
\ee
where $\langle M\rangle$ denotes the quadratic variation. Then
\begin{equation}\label{eq:Rxi}
R^{\bze}(t)=  \frac{(X(0))^p}{p} \mathcal{E}\left(\int_{0}^{t}(p\,\bzt(u) \bsi(u)+Z(u))\, d\bW(u)\right)\exp{\left(\int_{0}^{t}g(u,Z(u))\,du\right)},
\end{equation}
where
\be
g(u,z):= h(u,z)+\frac{{1}}{2}||z||^{2}+ p r + p \bzt(u) ( \bmu(u) + \bsi(u) z ) + \frac{p^{2}-p}{2}\norm{\bzt(u)\bsi(u)}^2.
\ee
Since $R^{\bzt}$ should be a supermartingale for every admissible $\bze(u)$ (and a martingale for some element $\bze^{*}(u)$), then $g$ has to be a non--positive process. With this in mind, a suitable candidate would be
\be
h(u,z):= -pr-\frac{{1}}{2}||z||^{2}+{\inf_{\bze(u)\in\mathcal{A}(u)}}\left\{-p\bzt(u) ( \bmu(u) + \bsi(u) z ) +\frac{p-p^{2}}{2}\norm{\bzt(u)\bsi(u)}^2 \right\},
\ee
which leads to

\begin{eqnarray}\label{BSDEf}
h(u,z)&=&-pr-\frac{{1}}{2}||z||^{2}+ \frac{p}{2(p-1)} \norm{\bsy{\sigma'}(u)(\bsi \bsy{\sigma'})^{-1}(u) (\bmu(u) + \bsi(u) z)}^2\\\notag &+& {\frac{p(1-p)}{2}}\mathrm{dist}\left(\frac{\bsy{\sigma'}(u)(\bsi \bsy{\sigma'})^{-1}(u) (\bmu(u) + {\bsi(u)} z)}{1-p}; \AA_u^\rho \bsi(u)\right)^2.
\end{eqnarray}
If in addition we let

\begin{equation}\label{BSDEf2}
\nonumber
\tilde{z}:= \frac{\bsy{\sigma'}(u)(\bsi \bsy{\sigma'})^{-1}(u) (\bmu(u) + \bsi(u) z)}{1-p} \quad \textrm{ and } \quad \tilde{\AA}_u^{\rho}:= \AA_u^{\rho} \bsi(u),
\end{equation}
then
\begin{small}
\be
\mathrm{dist}\left(\frac{\bsy{\sigma'}(u)(\bsi \bsy{\sigma'})^{-1}(u) (\bmu(u) + \bsi(u)z)}{1-p}; \AA_u^\rho \bsi(u)\right)^2=\|\frac{\bsy{\sigma'}(u)(\bsi \bsy{\sigma'})^{-1}(u)  (\bmu(u) + \bsi(u) z)}{1-p}-{\bze^{*}}'(u) \bsi(u)\|^2
\ee
\end{small}
with
\be
{\bze^{*}}'(u) \bsi(u)\in\mathrm{Proj(\tilde{Z}(u),\tilde{\AA}_u^{\rho})}.
\ee
The available results on existence of solutions to BSDEs require, to begin with, the predictability of the driver $h$. In our case this is closely related to the predictability of $\bze^{*},$ in other words, to whether or not the candidate for an optimal strategy is acceptable.

\begin{theorem}\label{th:opti} Let $Z$  be a predictable process such that
\be
\mathbb{E}\left(\int_0^T \norm{Z(u)}^2\, du\right)^{\frac{1}{2}}<\infty,
\ee
then for $(t, \omega)\in [0, T]\times\Omega,$ the mapping

\be
(t, \omega)\mapsto{\text{dist}}(\tilde{Z}_t(\omega), \tilde{\AA}_t^{\rho}(\omega)),
\ee
where $\tilde{Z}$ is as in Equation~\eqref{BSDEf2}, is predictable. In addition there exists a predictable process ${\bze}^{\ast}$ in $\R^n$ such that
\be
\mathbb{E}\left(\int_0^T \norm{{\bze^*}'(u) \bsi(u)}^2\, du \right)^{\frac{1}{2}}<\infty
\ee
and
\be \text{dist}\big(\tilde{Z}_t, \tilde{\AA}_t^{\rho}\big)=\text{dist}(\tilde{Z}_t,{\bze^{\ast}}'(t) \bsi(t)), \quad \forall t\in[0,T], \; P-a.s.. \ee
\end{theorem}
\noindent\begin{Proof} Let us define for $k\in\N$

\be
\AA_{t, k}^{\rho}(\omega):=\left\{\bzeta\in [-k, k]^n\,\Big|\, \rho_t(\e(\zeta,t))(\omega) - K_t(\omega) \le 0\right\}.
\ee
The purpose of artificially bounding the values of $\AA_{\cdot}^{\rho}$ is to make use of the theory of compact--valued correspondences (see Appendix~\ref{app:Corr}).
 It follows from Lemma~\ref{lemma:compact} that for all $k\in\N$ and for all $(t, \omega),$ the set ${\AA}_{t, k}^{\rho}(\omega)$ is non--empty and compact. Moreover, Proposition~\ref{prop:w-measurable} guarantees that for all $t\in [0, T]$ and $k\in\N,$ the correspondence $(\omega,t)\mapstto\tilde{\AA}_{t, k}^{\rho}(\omega)$ is weakly $\Pp$--measurable   (see Definition $A.2$ in the Appendix for the definition of weakly measurability).  Let $(C(\R^m), {\mathcal H})$ denote the space of non--empty, compact subsets of $\R^m,$ equipped with the Hausdorff metric. This is a complete, separable metric space, in which $\tilde{\AA}_{t, k}^{\rho}(\cdot)$  takes its values. Theorem~\ref{theo:Caratheodory-dist} then states that for $z\in\R^m$ and $t\in [0, T],$ the distance mapping

\be
\delta(\omega, z)={\text{dist}}\big(z, \,{\AA}_{t, k}^{\rho}(\omega)\bsi(t)\big)
\ee
is a Carath\'eodory one. Since the process $\tilde{Z}_t$ is predictable and $z\mapsto\delta(z, \omega)$ is continuous for all $\omega\in\Omega,$ the map

\be
(\omega,t)\mapsto{\textrm{dist}}\big(\tilde{Z}_t(\omega), \,{\AA}_{t, k}^{\rho}(\omega)\bsi(t)\big)
\ee
is $\Pp$--measurable. Finally

\be
{\text{dist}}\big(\tilde{Z}_t(\omega), \tilde{\AA}_t^{\rho}(\omega)\big)=\inf_{k\in\N}\big\{{\textrm{dist}}\big(\tilde{Z}_t(\omega), \,{\AA}_{t, k}^{\rho}(\omega)\bsi(t)\big)\big\},
\ee
thus the mapping $\omega\mapsto{\text{dist}}(\tilde{Z}_t(\omega), \tilde{\AA}_t^{\rho}(\omega))$ is predictable as the pointwise infimum of predictable ones. We now turn {our attention} to the second claim.
First we observe that since $\tilde{\AA}_t^{\rho}(\omega)$ is closed (and contained in $\R^m$), the set

\be
\overline{\AA}_t^{\rho}(\omega):={\textrm{argmin}}_{a\in\tilde{\AA}_t^{\rho}(\omega)}\Big\{{\text{dist}}(\tilde{Z}_t(\omega), a)\Big\}
\ee
is compact. It follows from the Measurable Maximum Theorem (\cite{AB06}, page 605)  that the correspondence $(t, \omega)\mapstto \overline{\AA}_t^{\rho}(\omega)$ is weakly $\Pp$--measurable. It is then implied by the Kuratowski--Ryll--Nardzewski Selection Theorem that $\overline{\AA}_{\cdot}^{\rho}(\cdot)$ admits a measurable selection ${{\bze}^{\ast}}' \bsi;$ in other words, there exists a predictable process ${\bze}^{\ast}:[0, T]\times\Omega\to\R^n$ such that

\be
{\textrm{dist}}(\tilde{Z}_t(\omega), \tilde{\AA}_t^{\rho}(\omega))={\textrm{dist}}(\tilde{Z}_t(\omega), {\bze}^{\ast}(t, \omega))\quad{\text{and}}\quad{{\bze}^{\ast}}'(t, \omega) \bsi(t,\omega)\in\tilde{\AA}_t^{\rho}(\omega).
\ee
Finally using the fact that the strategy $(0,\ldots,0)$ belongs to $\tilde{\AA}_\cdot^{\rho}$ we have that

\begin{eqnarray}
\label{eq:est1}
\int_0^T \norm{{\bze^*}'(u) \bsi(u)}^2 du &\leq& 2 \int_0^T \norm{{\bze^*}'(u) \bsi(u) - \tilde{Z}_u}^2 du + 2 \int_0^T \norm{\tilde{Z}_u}^2 du\nonumber\\
&=& 2 \int_0^T {\textrm{dist}}(\tilde{Z}_u, \tilde{\AA}_u^{\rho})^2 du + 2 \int_0^T \norm{\tilde{Z}_u}^2 du\nonumber\\
&\leq& 4 \int_0^T \norm{\tilde{Z}_u}^2 du <\infty.
\end{eqnarray}
\end{Proof}

\noindent To finalize, we must show that the quadratic--growth BSDE~\eqref{BSDE} admits a solution. In the following we will make use of the notion of BMO--martingale.
\begin{definition}
A continuous martingale $M$ is a BMO--martingale, if there exists a positive constant $a>0$ such that for every stopping time $\tau\leq T$,
\be
\E\left[ \langle M \rangle_T-\langle M \rangle_\tau\right] \leq a, \; \PP-a.s..
\ee
We will use the following property of BMO--martingales (which can be found in \cite{Kazamaki94}): if $M$ is a BMO--martingale then $\mathcal{E}(M)$ is a true martingale.
\end{definition}
\noindent We require the following result of Morlais \cite[Theorem 2.5 and Lemma 3.1]{Morlais09}, which extends the results
of Kobylanski~\cite{Ko00}:

\begin{theorem}\label{th:BSDE}
Let $h:[0,T]\times\Omega\times\mathbb{R}^{m}\rightarrow\mathbb{R}$ be measurable. Assume that there exist a predictable process $\alpha$ and positive constants $C_1, C_2$ satisfying $\alpha \geq 0$ and
\be
\int_0^T  \alpha_s ds \leq C_1, \; \PP-a.s..
\ee
If $h$ is such that
\begin{enumerate}
\item $z \mapsto h(u,z)$ is continuous
\item $|h(u, z)|\leq C_2\|z\|^2+\alpha_u,$
\end{enumerate}
then the BSDE \eqref{BSDE} with driver $h$ admits a solution $(Y,Z),$ where $Y$ and $Z$ are predictable processes with $Y$ bounded and $Z$ satisfying $\mathbb{E}\left(\int_0^T \|Z(t)\|^2 dt\right)^{\frac{1}{2}}<\infty$. In addition, the process $\int_0^\cdot Z(s) d\bW(s)$ is a BMO martingale and hence $\mathcal{E}\left(\int_0^\cdot Z(s) d\bW(s)\right)$ is a true martingale.
\end{theorem}
\noindent
The previous result allows us to show that the BSDE~\eqref{BSDE} with driver given by Equation~\eqref{BSDEf} admits a unique solution. Note that the fact that $\mathcal{E}\left(\int_0^\cdot Z(s) d\bW(s)\right)$ is a true martingale is essential in our approach since it basically allows the process $R^{\bze^\ast}$ to be a (true) martingale for some element $\bze^\ast$.
\begin{corollary}
\label{co:BSDE}
There exists a unique pair of predictable processes $(Y,Z)$ with $Y$ bounded and $Z$ satisfying $\mathbb{E}\left(\int_0^T \|Z(t)\|^2 dt\right)<\infty$ solution to the BSDE \eqref{BSDE} with driver given by Equation~\eqref{BSDEf}. In addition, the processes $\int_0^\cdot Z(s) d\bW(s)$ and $\int_0^\cdot {\bze^*}'(u) \bsi(u) d\bW(u)$ are BMO--martingales with $\bze^*$ given by Theorem \ref{th:opti}.
\end{corollary}
\begin{Proof} We apply Theorem~\ref{th:BSDE}, and measurability of $h$ is guaranteed by Theorem~\ref{th:opti}. The continuity in $z$ of the driver is straightforward, as are the growth conditions, given Assumption \ref{ass:boundmprx}. Again by Theorem~\ref{th:BSDE}, $\int_0^\cdot Z(s) d\bW(s)$ is a BMO--martingale which by definition, means that there exists a positive constant $a>0$ such that for every stopping time $\tau$,
$$ \E\left[ \int_\tau^T \norm{Z(s)}^2 ds\right] \leq a, \; \PP-a.s..$$
Hence, by Estimate~\eqref{eq:est1}, we have for any stopping time $\tau$ that
\be
\E\left[\int_\tau^T \norm{{\bze^*}'(u) \bsi(u)}^2 du \Big\vert \mathcal{F}_\tau\right] \leq 4 \E\left[\int_\tau^T \norm{\tilde{Z}_u}^2 du \Big\vert \mathcal{F}_\tau\right],
\ee
showing that $\int_0^\cdot {\bze^*}'(u) \bsi(u) d\bW(u)$ is a BMO--martingale since $\bsy{\sigma'}\bsy(\bsi \bsy{\sigma'})^{-1}\bsi \bmu$ is uniformly bounded by Assumption~\ref{ass:boundmprx}.

\end{Proof}

\noindent
We conclude with the existence of an optimal strategy to Problem \eqref{eq:opti}.
\begin{theorem}
Under the assumptions made above there exists an {acceptable} strategy $\bze^\ast$ {that solves}  Problem \eqref{eq:opti}. {If} we define the value function $v(x)$ as:
\be
v(x):= max_{\bze \in \AA^{\rho}} \mathbb{E}(U_{p}(\Xze(T))), \quad x>0
\ee
with $\AA^{\rho}$ the set of admissible $\R^n$-valued predictable processes $\bze$ such that $\bze(t)\in\AA^{\rho}_t$ for all $t$ in $[0,T]$ and $\Xze(0)=x,$ then it holds that
\be
v(x)=U_{p}(x) \exp(Y_0).
\ee
Here $(Y,Z)$ is a solution to the BSDE~\eqref{BSDE} with driver given by Equation~\eqref{BSDEf} and
\be
{\bze^{*}}'(u) \bsi(u)\in\mathrm{Proj\big(\tilde{Z}(u),\tilde{\AA}_u^{\rho}\big)}.
\ee
\end{theorem}
\begin{Proof} The existence of a solution to the BSDE \eqref{BSDE} is guaranteed by Corollary~\ref{co:BSDE}. Furthermore, the process
\be
\mathcal{E}\left(\int_{0}^{t}(p\,{\bze^{*}}'(u) \bsi(u)+Z(u))\, d\bW(u)\right)
\ee
is a true martingale since $\int_0^\cdot Z(s) d\bW(s)$ and $\int_0^\cdot {\bze^*}'(u) \bsi(u) d\bW(u)$ are BMO--martingales (with $\bze^*$ given as in Theorem \ref{th:opti}) by Corollary \ref{co:BSDE}. Now, as in \cite[Theorem 14]{HIM05}, for any admissible $\bzt$, the process $R^{\bzt}$ given by Equation~\eqref{eq:Rxi} is a supermartingale. Indeed, by construction $g$ is non-positive and the stochastic exponential $\mathcal{E}\left(\int_{0}^{t}(p\,\bzt(u) \bsi(u)+Z(u))\, d\bW(u)\right)$ is local martingale. Let $(\tau_n)_n$ be a localizing sequence associated to it. We have for every $n$ (and $s\leq t$) that: $\E[R_{t\wedge \tau_n}^{\bzt} \vert \mathcal{F}_s] \leq R_{s\wedge \tau_n}^{\bzt}$ and $R^{\bzt}$ is a non--negative process. Thus, Fatou's Lemma implies that
\be
\E[R_t^{\bzt} \vert \mathcal{F}_s] \leq R_s^{\bzt}.
\ee
Using the martingale optimality principle, we have that the processes {$R^{\bze}$} are well--defined and satisfy requirements (a) and (b). In addition, by construction, the processes $\bze^\ast$ such that $R^{\bze^\ast}$ {is a martingale} are those such that ${\bze^{*}}'(u) \bsi(u)\in\mathrm{Proj(\tilde{Z}(u),\tilde{\AA}_u^{\rho})}$.  Theorem \ref{th:opti} {yields} that these elements $\bze^\ast$ are admissible strategies, thus optimal. Take such an optimal strategy $\bze^*$. We have that
\be
 v(x)=\mathbb{E}(U_{p}(\Xoze(T))=\mathbb{E}(U_{p}(R^{{\bze}^\ast}(T))=R^{\bze^\ast}(0)=U_{p}(x) \exp(Y_0).
 \ee

\end{Proof}

\noindent
The previous result admits a dynamic version:
\begin{theorem}
\label{th:dynamicvalue}
Let $v(t,x)$ be the dynamic value function defined as:
\be
v(t,x):=\esssup_{\bze \in \AA^t} \mathbb{E}\left(U_p\left(x+\int_t^T \bze(s) X^{\bze}_s \frac{dS_s}{S_s}\right)\Big\vert \mathcal{F}_t\right) \quad t\in[0,T], \; x>0,
\ee
where $\mathcal{A}^t:=\{\bze \in \AA^\rho, \; \bze(s)=0, \; s<t\}$.
Then
\be
v(t,x)=U_p(x) \exp(Y_t),
\ee
where $(Y,Z)$ is a solution to the BSDE \eqref{BSDE} with driver given by Equation \eqref{BSDEf} and
\be
{\bze^{*}}'(u) \bsi(u)\in\mathrm{Proj(\tilde{Z}(u),\tilde{\AA}_u^{\rho})}.
\ee
\end{theorem}

\begin{Proof} Let $\bze$ any element of $\AA$ and $\bze^\ast$ such that the associated $R^{\bze^\ast}$ is a martingale. Then by definition of the $R^{\bze}$ processes, we have that $R^{\bze}(t)=U_p(x) \exp(Y_t)$ since $\bze(s)=0$ for $s<t$ and so
\begin{eqnarray*}
&&\mathbb{E}\left(U_p \left(x+\int_t^T \bze(s) X^{\bze}_s \frac{dS_s}{S_s}\right) \Big\vert \mathcal{F}_t\right)\\
&=& \mathbb{E}\left(R^{\bze}(T)\vert \mathcal{F}_t\right)\\
&\leq& R^{\bze}(t)=U_p(x) \exp(Y_t)=\mathbb{E}\left(R^{{\bze}^\ast}(T)\vert \mathcal{F}_t\right)=\mathbb{E}\left(U_p \left(x+\int_t^T {\bze}^\ast(s) X^{\bze^\ast}_s \frac{dS_s}{S_s}\right) \Big\vert \mathcal{F}_t\right).
\end{eqnarray*}
Hence, $v(t,x)=U_p(x) \exp(Y_t)$.

\end{Proof}

\begin{remark}
Sometimes one {might be} interested in another version of the dynamic value function above. Given an element $\bze$ in $\AA^\rho$ {they may} consider the quantity
\be
 v(t,X_t^{\bze}):=\esssup_{\tilde{\bze} \in \mathcal{A}^{t,\bze}} \mathbb{E}\left(U_p\left(X_t^{\bze}+\int_t^T \tilde{\bze}(s) X^{\tilde{\bze}}_s \frac{dS_s}{S_s}\right)\Big\vert \mathcal{F}_t\right), \quad t\in[0,T],
\ee
where
$\mathcal{A}^{t,\bze}:=\{\tilde{\bze} \in \AA^\rho, \; \tilde{\bze}(s)=\bze(s), \; s\leq t\}$. Then we have that
$v(t,X_t^{\bze})=U_p(X_t^{\bze}) \exp(Y_t)$ where $(Y,Z)$ is the unique solution of the BSDE \eqref{BSDE} with driver given by Equation \eqref{BSDEf}.
\end{remark}

\begin{remark}
The stochastic process $\exp(Y_t)$ in the expression of the value function is sometimes called the \textit{opportunity process}, since it gives the value of the optimal wealth with initial capital one unit of currency (see \cite{Nutz}).
\end{remark}

\begin{remark}
Notice that for the sake of the explanation, we have chosen to fix the risk aversion coefficient $p$ in $(0,1)$ but we can also consider the case where $p<0$. Then the driver $h$ given by Equation \eqref{BSDEf} has to be modified suitably.
\end{remark}

\section{Time Consistent Distortion Risk Measures}\label{sec:VaR}

In this section we define a broad class of families of risk measures that are \textit{time consistent}. We show that, under the constrains imposed by members of this class, optimal investment strategies follow a three--fund separation behavior. Let
\be
\rho_t(\e(\bzeta,t)-1)(\omega) := \exp\Big\{r\tau\Big\} \e_1(\bzeta, t)(\omega)\rho_0\big(\e_2(\bzeta, t)(\omega)-1\big),
\ee
where
\be
\e_1(\bzeta, t)(\omega):= \exp\Big\{\tau\big(\bzeta x-\frac{1}{2}\norm{\bzeta y}^2\big)\Big\}\Big|_{x=\bmu(\omega, t), y=\bsi(\omega, t)}
\ee
and
\be
\rho_0\big(\e_2(\bzeta, t)(\omega)-1\big):= \rho_0\Big(\exp\big\{x y\Delta_{\tau}W_0\big\}-1\Big)\Big|_{x=\bzeta(\omega, t), y=\bsi(\omega, t)}.
\ee
Here  $ \rho_0$ a distortion risk measure, i.e.
\be
\rho_0 (X) = \int _{[0,1]}  F^{-1}_{X} (u) dD(u),
\ee
where $F^{-1}_{X}$ is the inverse CDF of $X,$ and $D$ is a distortion,
i.e., it is right--continuous, increasing on $[0,1],$ $D(0)=0$ and $D(1)=1.$
The choice $D(u)={\bf 1}_{\{ u\geq 1-\alpha\}}$ yields VaR$_{\alpha}$ and $D(u)=\frac{1}{\alpha} [u-(1-\alpha)]^{+}$ yields TVaR$_{\alpha}.$ LEL$_{\alpha}$ can be recovered by choosing $D(u)=\frac{1}{\alpha} [u-(1-\alpha)]^{+} {\bf 1}_{\bmu=0}$ (since LEL$_{\alpha}$ is TVar$_{\alpha}$ computed under one of the risk neutral probability measures). Distortion risk measures form a rich class, which contains: proportional hazards, proportional odds, Wang transform,
positive Poisson mixture, etc. It follows from direct computations that
\be
\rho_t(\e(\bzeta,t)-1)= \int _{[0,1]}   \left[
      1-\exp\Big(\tilde{Q}(\zmu(t),\zsigma(t))\tau+N^{-1}(u)\zsigma(t)\sqrt{\tau}\Big)
      \right] dD(u).
\ee
In the light of this, one can see that Assumption \eqref{ass:rho} holds true. From this point on we work under the assumption that $K_t <1.$ This implies (quite naturally) that the risk should be smaller than the current position.

\subsection{A common form of the risk constraints}

Below we present some properties of the constraint sets $\AA^{\rho}.$
\begin{proposition}\label{pro:common-form}
Each constraint set $\AA^{\rho}_{t}$ can be expressed as
\begin{equation}%
\label{equ:representation-common-form}
    \nonumber
    \begin{split}
\AA^{\rho}_{t}= \sets{\bze\in\R^m}{f(\bzt\bmu(t), \norm{\bzt\bsi(t)})\leq {K}_t},
    \end{split}
\end{equation}
for some function $f:\R\times [0,\infty)\to\R\cup\set{\infty},$ which satisfies
\be
f \in C^1(\R\times [0,\infty)),\quad f (0,0)\leq 0,\quad  \lim_{\zeta\rightarrow \infty} f(\bzt\bmu(t), \norm{\bzt\bsi(t)})=1.
\ee
\end{proposition}
\begin{Proof} The function $f$ is defined by
\be
f(x,y)= \int _{[0,1]}   \left[
      1-\exp\Big((r+x-\frac{y^2}{2})\tau+N^{-1}(u)y\sqrt{\tau}\Big)
      \right] dD(u),
\ee
so it follows that $f \in C^1(\R\times [0,\infty)), f (0,0)\leq 0.$ In the light of
\be \lim_{\zeta\rightarrow \infty} \left[
      1-\exp\Big(\tilde{Q}(\zmu(t),\zsigma(t))\tau+N^{-1}(u)\zsigma(t)\sqrt{\tau}\Big)
      \right] =1,
\ee
it follows that $ \lim_{\zeta\rightarrow \infty} f(\bzt\bmu(t), \norm{\bzt\bsi(t)})=1.$

\end{Proof}

\noindent The choice of the threshold ${K}_t<1$ and Proposition~\ref{pro:common-form} yield the compactness of the constraint sets
associated with the risk measures considered in this section.

\subsection{A Three-Fund Separation Result}

In this section we further characterize the optimal investment strategy. Let us recall that  $\bze^{*}$ it is given by
  $${\bze^{*}}'(u) \bsi(u)\in\mathrm{Proj(\tilde{Z}(u),\tilde{\AA}_u^{\rho}),\quad u\in [0,T]}.$$
  Compactness of $\AA^{\rho}$ leads to compactness of $\tilde{\AA}^{\rho}$ which in turn yields the
  existence of the projection.

\begin{theorem}\label{L:optimal_policy}
 There exist two stochastic processes $\beta^{*}_{1}$ and $\beta^{*}_{2}$ such that optimal strategy $\bze^{*}$
 can be decomposed as
\begin{equation}\label{optima1}
\bze^{*}(t)=\frac{\beta^{*}_{1}(t)}{1-p}\bze_{M}(t)+\beta^{*}_{2}(t)(\bsi(t)\bsi^{'}(t))^{-1}\bsi(t)Z(t),\quad 0\leq t\leq T,
\end{equation}
where $Z(t), 0\leq t\leq T$ is part of the $(Y,Z)$ solution of BSDE \eqref{BSDE} with driver \eqref{BSDEf}.

\end{theorem}

\begin{Proof} We cover the case $p\geq 0$ only (the $p<0$ case can be obtained by an analogous argument). Let recall that for a fixed path $\omega,$ the optimal strategy $\bze^{*}(t)$ solves
\be
\bze^{*}(t)=\arg\min_{\bze\in\mathcal{A}(t)}\left\{-p\bzt ( \bmu(t) +p \bsi(t) Z(t) ) + \frac{p-p^{2}}{2}\norm{\bzt\bsi(t)}^2 \right\}.
\ee
The convex, quadratic functional
\be
\bze\rightarrow H(t,\bze):= -p\bzt ( \bmu(t) +p \bsi(t) Z(t) ) + \frac{p-p^{2}}{2}\norm{\bzt\bsi(t)}^2
\ee
 is minimized over the constraint set $\mathcal{A}(t)$ at a point $\bze^{*}(t).$
 which is either an absolute minimum or else should be on the boundary of $\mathcal{A}(t).$ Thus, for a fixed path,  $\bze^{*}(t)$ minimizes
 $H(t,\bze)$ over the constraint $f (\bzt\bmu(t), \norm{\bzt \bsi(t)})\leq \hat{K}_t.$ The solution $\bze^{*}(t)$ is not the zero vector, since the zero vector is not an absolute minimum and $f(0,0)\leq0.$ For $\bze\neq0$, it follows that
\be
\nabla f(\bzt\bmu(t), \norm{\bzt \bsi(t)})=  f_{1}(\bzt\bmu(t), \norm{\bzt \bsi(t)})  \bmu(t) -\frac{ f_{2}(\bzt\bmu(t), \norm{\bzt \bsi(t)})  }{||\bzt \bsi(t)||}\bsi(t)\bsi{'}(t)\bze,
\ee
where $f_{1}$ and $f_{2}$ stand for the partial derivatives of function $f$. According to the \textrm{Karush--Kuhn Tucker} Theorem,
either $\nabla f(\bzt\bmu(t), \norm{\bzt \bsi(t)})=0$ or else there is a positive
$\lambda$ such that \begin{equation}\label{grad} \nabla
  H(t,\bze) =\lambda \nabla f(\bzt\bmu(t), \norm{\bzt \bsi(t)}).\end{equation} In both cases, straightforward computations, show that $\bze^{*}(t)$ should have the form given in Equation~\eqref{optima1}.

\end{Proof}

Theorem~\ref{L:optimal_policy} is a three-fund separation result. It states that a utility--maximizing investor who is subject to
regulatory constraints will invest his wealth into three-funds: 1. the savings account; 2. a risky fund with
return $\bze_{M}(t), t\in [0, T];$ 3. a risky fund with return $(\bsi(t)\bsi'(t))^{-1}\bsi'(t)Z(t) , t\in [0, T].$ Most of
the results in the financial literature are two--funds separation ones (optimal wealth being invested into
a saving account and a risky fund). We would obtain such a  two--funds separation result if we restricted our model to one in
which stocks returns and volatilities were deterministic. It is a consequence of the randomness of
the stocks returns and volatilities that the optimal investment includes an extra risky fund. Investment in the latter fund
can be regarded as a hedge against risk implied by stochastic stock returns and volatilities.

\begin{remark}
For the special case of $\rho_0=$TVaR$_\alpha$ the associated acceptance set  $\AA^{\rho}$ is convex; this is
also the case when  $\rho_0=$VaR$_\alpha,$ whenever $\alpha\in[0,0.5].$
The convexity of  $\AA^{\rho}$  implies the uniqueness of optimal trading strategy $\bze^{*},$ a fact that turns out to be useful in numerical implementations.
\end{remark}

\section{A numerically implemented example}\label{section:numerics}

In this section we present numerical simulations for the constrained optimal strategies and   the associated constrained \textit{opportunity processes}. Recall that by opportunity process we mean the process $\exp(Y_t),$ which appears in the value function $v(t,x)$ in Theorem \ref{th:dynamicvalue}; that is $v(t,x)=\frac{x^p}{p} \exp(Y_t)$. The opportunity process represents the value function of an investor with initial capital one dollar. It is a stochastic process and in the figures below we present one sample path. For simplicity and the numerical tractability of the analysis we assume that we deal with one risky asset ($n=1$), one bond with rate zero ($r=0$) and one Brownian motion ($m=1$). In addition, we assume that the risky asset is given by the following SDE:
\be
dS_t=S_t(\textbf{1}_{[-1,1]}	(W_t) dt + dW_t), \quad t\in [0,1] \quad (T=1), \quad S_0=1.
\ee
Our simulation relies on numerical schemes for quadratic growth BSDEs. We use the scheme of Dos Reis and Imkeller~\cite{DosReisImkeller,DosReis}. The latter, in a nutshell, relies on a truncation argument of the driver, and it reduces the numerical--simulation problem to one of a BSDE with a Lipschitz--growth driver . Here we use the so--called \textit{forward scheme} of Bender and Denk~\cite{BenderDenk}.

\begin{figure}[!hbt]
  \includegraphics[width=12cm]{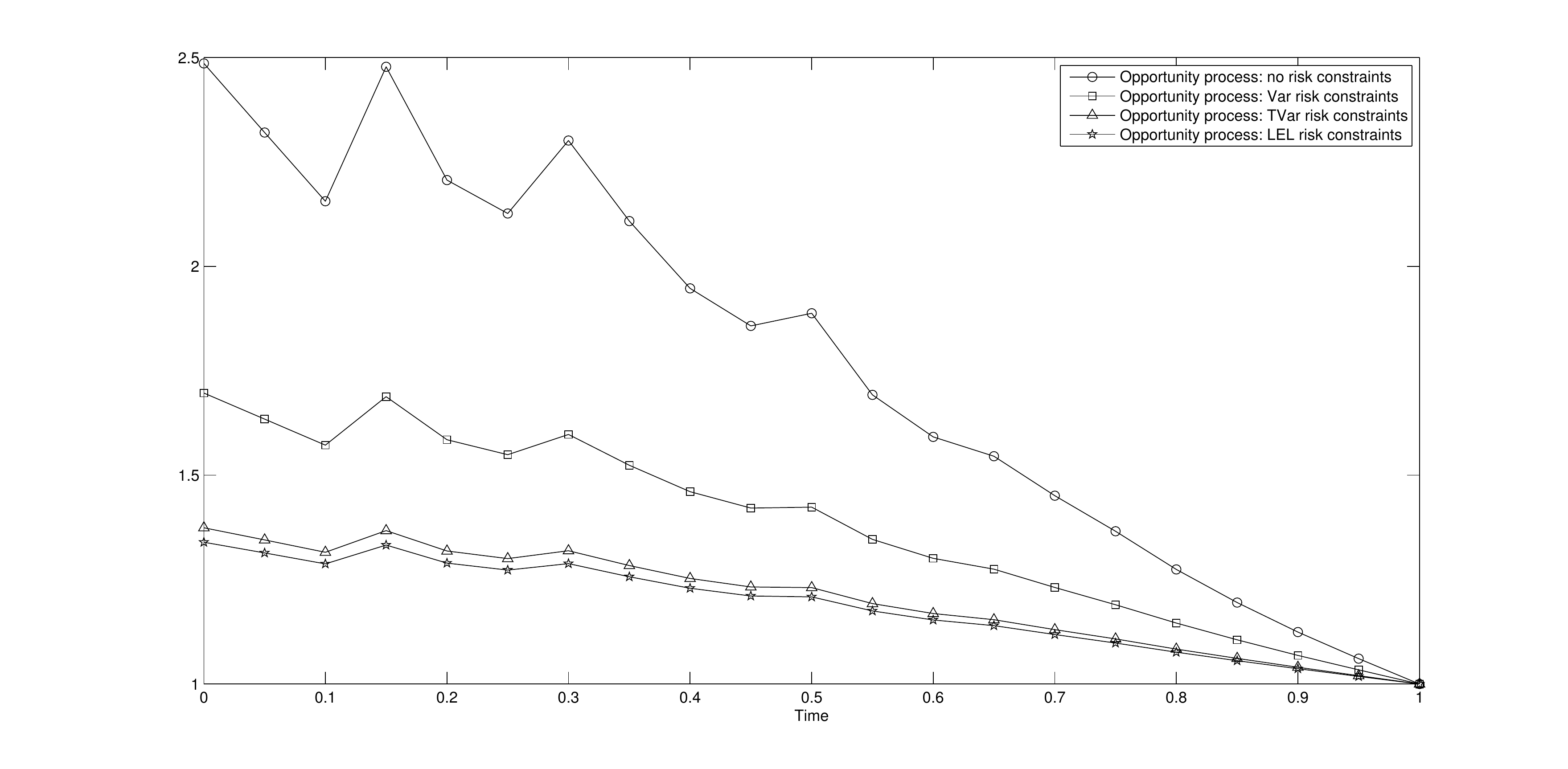}\\
  \caption{Constrained and unconstrained opportunity processes.}\label{fig:Opport}
\end{figure}

\begin{figure}[!hbt]
  \includegraphics[width=12cm]{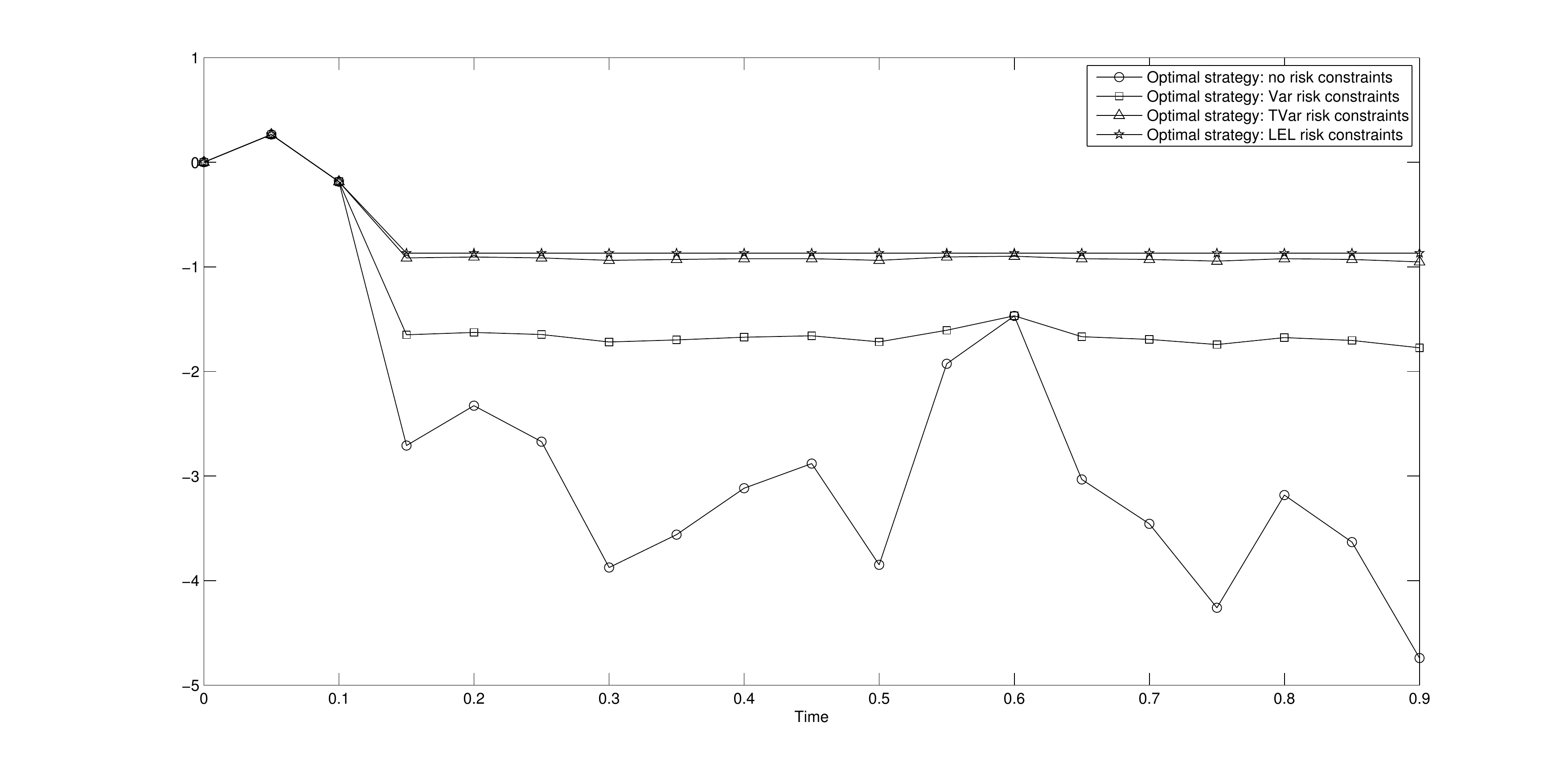}\\
  \caption{Constrained and unconstrained optimal strategies.}\label{fig:Strat}
\end{figure}

In Figure~\ref{fig:Opport} we illustrate the opportunity processes arising from imposing VaR , TVar and LEL.  We have used the following set of parameters: $p$=\texttt{0.85}, $\alpha$=\texttt{0.10}, $K$=\texttt{0.3} and $T=\texttt{1}$. The time discretization is $1/15$  and $\tau$=\texttt{1/15}. The unconstrained opportunity process is also presented. The corresponding trading strategies are presented in Figure~\ref{fig:Strat}. We observe a spike in the opportunity process that may be explained by gambling; indeed looking at the TVaR constrained optimal strategy we see
that it differs considerably from the unconstrained one (in which the stock is shorted). This finding supports the idea that risk constraints reduce speculation.

\section{Conclusions}

We have analyzed, within an incomplete--market framework, the portfolio--choice problem of a risk averse agent (who is characterized by CRRA preferences), when risk constraints are imposed continuously throughout the investment phase. Using BSDE technology, in the spirit of~\cite{HIM05}, has enabled us to allow for a broad range of risk measures that give rise to the risk constraints, the latter being (possibly) time--dependent. In order to use such technology, we have made use of Measurable Selections theory, specifically when addressing the issue of the driver of the BSDE at hand. We have characterized the optimal (constrained) investment strategies, and in the case of distortion risk measures we have provided explicit  expressions for them. Here we have shown that optimal strategies may be described as investments in three funds, which is in contrast with the classical two--fund separation theorems. Finally, using recent results in~\cite{DosReisImkeller}, we have provided some examples that showcase the way in which our dynamic risk constraints limit investment strategies and impact utility at maturity.

\appendix

\section{Properties of the constraint sets $\AA_t^{\rho}$}\label{app:Corr}

Several analytical properties of the (instantaneous) constraint sets
$\AA_t^{\rho}$ are established in this section.   The analysis requires some core concepts of the theory of measurable correspondences\footnote{For a comprehensive overview of the theory of measurable correspondences, we refer the reader to~\cite{AB06}.}. We require the following auxiliary correspondences:
\be
\AA_{t, k}^{\rho}(\omega):=\left\{\bzeta\in [-k, k]^n\,\Big|\, \rho_t(\e(\bzeta,t)-1)(\omega) - K_t(\omega) \le 0\right\},\quad k\in\N.
\ee
The purpose of artificially bounding the values of $\AA_{\cdot}^{\rho}$ is to make use of the theory of compact--valued correspondences, which exhibit many desirable properties.
\begin{lemma}\label{lemma:compact}
For any $m\in\N,$ the correspondence $\AA_{\cdot, k}^{\rho}:[0, T]\times\Omega\twoheadrightarrow\R^n$ is non--empty and compact valued for almost all $(t, \omega)\in[0, T]\times\Omega.$
\end{lemma}
\noindent \begin{Proof}
The non--vacuity follows from the fact that $\bzeta\equiv 0,$ i.e. no wealth invested in risky assets, is an acceptable position. To show closedness of the sets $\AA_{t, k}^{\rho}(\omega),$ fix $\omega\in\Omega$ {and} consider a sequence $\big\{\bzeta_n\big\}\subset\AA_{t, k}^{\rho}(\omega)$ such that $\bzeta_n\to\bzeta.$
Using Assumption \ref{ass:rho} it holds that
$$ \rho_t(\e(\bzeta,t)-1)(\omega) - K_t(\omega) = \lim_{n\to \infty} \rho_t(\e(\bzeta_n, t)-1)(\omega) - K_t(\omega) \leq 0$$
holds for all $t\in [0, T]$ and
which implies that $\bzeta\in\AA_t(\omega).$ The latter, together with the fact that $\bzeta\in[-k, k]^n$ finalizes the proof.

\end{Proof}

\begin{definition}
A correspondence $\phi$ between a measurable space $(\Theta, \mathcal{G})$ and a topological space $X$ is said to be weakly measurable if for all $F\subset X$ closed, the lower inverse of $F,$ defined as
\be
\phi^l(F):=\left\{\theta\in\Theta\,\mid\, \phi(\theta)\cap F\neq\emptyset\right\},
\ee
belongs to $\mathcal{G}.$
\end{definition}
\noindent In the case of compact--valued correspondences, weak--measurability and Borel measurability (in terms of the Borel $\sigma$--algebra generated by the Hausdorff metric) are equivalent notions. Given a correspondence $\phi:\Omega\times [0,T]\mapstto\R^n$ we define the corresponding closure correspondence via $\bar\phi(\omega,t):=\overline{\phi(\omega,t)}.$ For notational purposes let
\be
f\big((t, \omega) , \bzeta\big) = \rho_t(\e(\bzeta,t)-1)(\omega) - K_t(\omega).
\ee
Recall that $\Pp$ denotes the predictable $\sigma$--algebra on $[0, T]\times\Omega.$ The function $f\big((\cdot, \cdot), \cdot\big)$ is a Carath\'eodory function with respect to $\Pp,$ i.e.
it is continuous in $\bzeta$ and $\Pp$--measurable in $(t, \omega).$

\begin{proposition}\label{prop:w-measurable}
For any $k\in\N,$ the correspondence $\AA_{\cdot, k}^{\rho}:[0, T]\times\Omega\twoheadrightarrow\R^n$ is weakly\\ $\Pp$--measurable.
\end{proposition}
\noindent\begin{Proof} Let $F\subset\R^n$ be closed and consider $\big\{\bzeta_m\big\}_{m=1}^{\infty}\subset F$ dense. For $\eta\in\N$ let
\be
^{\eta}\AA_{t, k}^{\rho}(\omega):=\left\{\bzeta\in [-k, k]^n\,\Big|\, f\big((t, \omega) , \bzeta\big) < \frac{1}{\eta}\right\}.
\ee
We have that
\begin{eqnarray*}
\big(\,^{\eta}\AA_{\cdot, k}^{\rho}\big)^l(F) & = & \Big\{ (t, \omega)\in [0, T]\times\Omega\,\big|\,f\big((t, \omega) , \bzeta\big)<\frac{1}{\eta}\quad{\textrm{for some}}\quad\bzeta\in F\Big\}\\
        & = & \Big\{ (t, \omega)\in [0, T]\times\Omega\,\big|\,f\big((t, \omega) , \bzeta_m\big)<\frac{1}{\eta}\quad{\textrm{for some}}\quad m\in\N\Big\}\\
        & = & \bigcup_{m=1}^{\infty} f^{-1}\big((\cdot, \cdot), \bzeta\big)\big(-\infty, \frac{1}{\eta}\big).
\end{eqnarray*}
The second equality holds because $f$ is continuous in $\bzeta,$ $\big\{\bzeta_m\big\}_{l=1}^{\infty}$ is dense and $(\infty, 1/\eta)$ is open. Since $f$ is Carath\'eodory, then $f^{-1}\big((\cdot, \cdot) , \bzeta\big)\big(-\infty, \frac{1}{\eta}\big)\in\Pp,$ hence for all $\eta\in\N,$ the correspondence $^{\eta}\AA_{\cdot, k}^{\rho}$ is weakly $\Pp$--measurable. Next we have
\be
\AA_{t, k}^{\rho}(\omega)\subset\overline{\,^{\eta}\AA_{t, k}^{\rho}(\omega)}\subset\left\{\bzeta\in [-k, k]^n\,\Big|\, f\big((t, \omega), \bzeta\big) \le \frac{1}{\eta}\right\},
\ee
where the second inclusion follows again from the continuity of $f$ in $\bzeta.$ This implies that
\be
\AA_{t, k}^{\rho}(\omega)=\bigcap_{\eta=1}^{\infty}\overline{\,^{\eta}\AA_{t, k}^{\rho}}(\omega),
\ee
and
\be
{\textrm{graph}}\big(\AA_{\cdot, k}^{\rho}(\cdot)\big)=\bigcap_{\eta=1}^{\infty}{\textrm{graph}}\big(\overline{\,^{\eta}\AA_{\cdot, k}^{\rho}}(\cdot)\big).
\ee
The graph of the closure of a weakly--measurable correspondence is measurable, hence ${\textrm{graph}}\big(\AA_{t, k}^{\rho}\big)$ is measurable, by virtue of being the (denumerable) intersection of measurable graphs. Since a compact--valued correspondence with a measurable graph is itself weakly--measurable (see Lemma 18.4  (part 3) and Corollary 18.8 in~\cite{AB06}), we conclude that the correspondence $(t, \omega)\mapstto \AA_{t, k}^{\rho}(\omega)$ has such property.

\end{Proof}

\noindent The following theorem, whose proof can be found in~\cite{AB06}, page 595, plays an important role in the proof of predictability of our BSDE's driver:

\begin{theorem}\label{theo:Caratheodory-dist}
A nonempty--valued correspondence mapping a measurable space into a separable, metrizable
space is weakly--measurable if and only if its associated distance function is a Carath\'eodory function.
\end{theorem}

\def\cprime{$'$} \def\cprime{$'$}
\providecommand{\bysame}{\leavevmode\hbox to3em{\hrulefill}\thinspace}
\providecommand{\MR}{\relax\ifhmode\unskip\space\fi MR }
\providecommand{\MRhref}[2]{%
  \href{http://www.ams.org/mathscinet-getitem?mr=#1}{#2}
}
\providecommand{\href}[2]{#2}

\end{document}